\documentclass[twocolumn,showpacs,preprintnumbers,amsmath,amssymb]{revtex4}

\usepackage{graphicx}
\usepackage{dcolumn}
\usepackage{bm}
\usepackage{color}

\begin{document}

\preprint{APS/123-QED}

\title{Critical behavior of magnetization in URhAl:\\Quasi-two-dimensional Ising system with long-range interactions\footnote{Phys. Rev. B {\bf 97}, 064423 (2018)}}

\author{Naoyuki Tateiwa$^{1}$}
\email{tateiwa.naoyuki@jaea.go.jp} 
\author{Ji\v{r}{\'i} Posp{\'i}\v{s}il$^{1,2}$}
\author{Yoshinori Haga$^{1}$}%
\author{Etsuji Yamamoto$^{1}$}%

\affiliation{
$^{1}$Advanced Science Research Center, Japan Atomic Energy Agency, Tokai, Naka, Ibaraki 319-1195, Japan\\
$^2$Charles University, Faculty of Mathematics and Physics, Department of Condensed Matter Physics, Ke Karlovu 5, 121 16 Prague 2, Czechia\\
}
\date{\today}

\begin{abstract}
 The critical behavior of dc magnetization in the uranium ferromagnet URhAl with the hexagonal ZrNiAl-type crystal structure has been studied around the ferromagnetic transition temperature $T_{\rm C}$. The critical exponent $\beta$ for the temperature dependence of the spontaneous magnetization below $T_{\rm C}$, $\gamma$ for the magnetic susceptibility, and $\delta$ for the magnetic isotherm at $T_{\rm C}$ have been obtained with a modified Arrott plot, a Kouvel-Fisher plot, the critical isotherm analysis and the scaling analysis. We have determined the critical exponents as $\beta$ = 0.287 $\pm$ 0.005, $\gamma$ = 1.47 $\pm$ 0.02, and $\delta$ = 6.08 $\pm$ 0.04 by the scaling analysis and the critical isotherm analysis. These critical exponents satisfy the Widom scaling law ${\delta}{\,}={\,}1+{\,}{\gamma}/{\beta}$. URhAl has strong uniaxial magnetic anisotropy, similar to its isostructural UCoAl that has been regarded as a three-dimensional (3D) Ising system in previous studies. However, the universality class of the critical phenomenon in URhAl does not belong to the 3D Ising model ($\beta$ = 0.325, $\gamma$ = 1.241, and $\delta$ = 4.82) with short-range exchange interactions between magnetic moments. The determined exponents can be explained with the results of the renormalization group approach for a two-dimensional (2D) Ising system coupled with long-range interactions decaying as $J(r){\,}{\sim}{\,}r^{-(d+{\sigma})}$ with $\sigma$ = 1.44. We suggest that the strong hybridization between the uranium $5f$ and rhodium $4d$ electrons in the U-Rh$_{\rm I}$ layer in the hexagonal crystal structure is a source of the low dimensional magnetic property. The present result is contrary to current understandings of the physical properties in a series of isostructural UTX uranium ferromagnets (T: transition metals, X: $p$-block elements) based on the 3D Ising model. 
 \end{abstract}


\maketitle

\section{Introduction}
\subsection{General introduction}
 Itinerant ferromagnets have attracted much attention because of their interesting physical properties, for example, unconventional superconductivity, exotic magnetic states such as skyrmion lattice or quantum critical behavior\cite{saxena,huxley1,aoki0,huy,robler,yu,smith,grigera,pfleiderer1,niklowitz,brando1}. In particular, many experimental and theoretical studies have looked at  novel phenomena related to a quantum phase transition (QPT) between ferromagnetic and paramagnetic states that can be tuned by external pressure, magnetic field or alloying constituent elements. 
 
 The unique features of actinide $5f$ systems is the co-existence of ferromagnetism and superconductivity that has been found in the uranium ferromagnetic superconductors UGe$_2$, URhGe, and UCoGe\cite{saxena,huxley1,aoki0,huy}. Superconductivity appears around the pressure-induced phase boundary of the ferromagnetism in UGe$_2$ and UCoGe\cite{saxena,huxley1,bastien}. URhGe shows ferromagnetic and superconducting transitions at $T_{\rm C}$ = 9.5 K and $T_{sc}$ = 0.25 K, respectively\cite{aoki0}. When magnetic field is applied along the magnetic hard $b$-axis in this orthorhombic crystal structure, field-induced reentrant superconductivity appears around $H_{\rm R}{\,}{\sim}{\,} 12$ T where the ferromagnetic transition temperature $T_{\rm C}$ is suppressed\cite{levy1,levy2}.
 
 Novel features of the physical properties under high pressure and high magnetic field have been extensively studied for the ferromagnetic superconductors UGe$_2$, URhGe, and UCoGe, and strongly uniaxial ferromagnets UCoAl, Ru-doped UCoAl, URhAl, and UCoGa \cite{taufour,nakamura,slooten,aoki,pospisil1,opletal,shimizu2,misek}. The line of continuous ferromagnetic transitions form a ``wing structure" in the temperature-pressure-magnetic field phase diagram of the uranium ferromagnets. When the pressure is applied, the paramagnetic to ferromagnetic transition changes from a second order to a first order transition at a tricritical point (TCP) before the critical pressure of the ferromagnetic state and the line bifurcates into finite magnetic fields at the TCP. Review papers gives the current status of experimental and theoretical studies on this subject\cite{belitz,brando2}. Generally, the ferromagnetic states in the uranium ferromagnets are strongly uniaxial. The experimental data have been discussed with theories based on the 3D Ising model.   

 The study of the critical behavior of the magnetization provides crucial information as to the type of the magnetic phase transition and nature of spin-spin interactions. We have found that the universality class of the critical phenomena in the uranium ferromagnetic superconductors UGe$_2$ and URhGe do not belong to any known universality classes of critical phenomena such as the 3D Ising model\cite{tateiwa1}. We suggest that uniaxial uranium ferromagnets have special features that cannot be understood only with the 3D Ising model. 
 
 In this paper, we report the critical behavior of the magnetization in URhAl. The low-dimensionality of the magnetism in URhAl is suggested by the analysis of the critical behavior using renormalization group theory. Recently, several studies have reported on the low-dimensionality of the ferromagnetism in the $3d$ electrons system Y$_2$Ni$_7$, Cr$_{11}$Ge$_{19}$, CrSiTe$_3$, CrGeTe$_3$, and Cr$_{0.62}$Te\cite{bhattacharyya,han,liu,liu2,lin,liu3}. Low-dimensionality of the magnetism has been rarely recognized in experimental studies of uranium intermetallics. We propose a view for this research field. 

  \subsection{Basic physical properties in URhAl}
 Two groups in uranium ferromagnets have been extensively studied from the viewpoint of the quantum phase transition between ferromagnetism and paramagnetism. One is the uranium ferromagnetic superconductors: binary orthorhombic UGe$_2$, URhGe and UCoGe with the orthorhombic TiNiSi-type structure. The other is a series of UTX uranium ferromagnets with the hexagonal ZrNiAl-type crystal structure, where T is a transition $d$ metal and X a $p$-block element\cite{sechovsky}. Among the latter UTX systems, UCoAl has been most studied for three decades\cite{aoki,sechovsky,matsuda1,javorsky1,kucera,matsuda2,nohara,karube1,matsuda3,morales,combier1,takeda,karube2,shimizu1,kimura,combier2}. The compound is a heavy fermion paramagnet and shows a metamagnetic transition at ${{\mu}_0}{H_m}{\,}{\sim}{\,}0.7 $ T at low temperatures when the magnetic field is applied along the magnetic easy $c$-direction. This metamagnetic transition terminates at a finite temperature critical end point (CEP) at $T_{\rm CEP}{\,}{\sim}{\,} 11$ K and ${\sim}$ 1 T. The magnetization and magnetic susceptibility show strong uniaxial anisotropy. Experimental data in UCoAl have been discussed from the viewpoint of quantum criticality in the 3D Ising system. In this study, we suggest that iso-structural URhAl should be regarded as a quasi 2D Ising system with long range magnetic interactions.

 We summarize the basic physical properties in URhAl. Figure 1 shows (a) the hexagonal ZrNiAl-type crystal structure ($P{\bar 6}2m$) of URhAl, (b) U-Rh$_{\rm {I}}$ and (c) Rh$_{\rm {II}}$-Al layers viewed along the $c$-axis. The structure is an ordered ternary derivative of the Fe$_2$P-type structure. There are two Rh sites in the structure in contrast to the uranium ferromagnetic superconductors URhGe or UCoGe crystalizing in the TiNiSi-type structure where only one transition metal site is present. One third of them lies on the U-Rh$_{\rm I}$ layer and the others on the Rh$_{\rm {II}}$-Al layer. The layers alternate with each other. Lattice parameters are $a$ = 0.69958 nm and $c$ = 0.400241 nm at room temperature\cite{paixao1}. The in-plane U-U distance (0.363 nm) is smaller than that along the $c$-direction that is equal to the lattice parameter $c$.
       \begin{figure}[t]
\includegraphics[width=8.5cm]{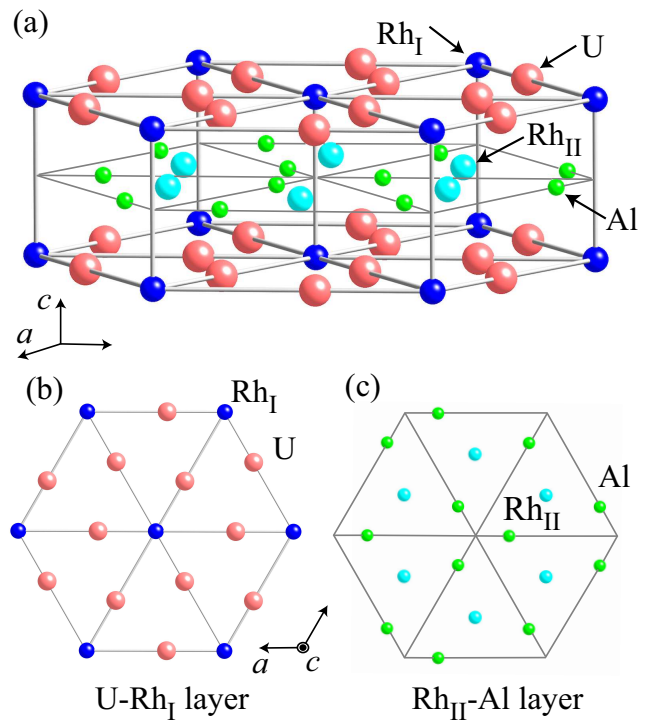}
\caption{\label{fig:epsart}(a)Representation of the hexagonal ZrNiAl-type crystal structure of URhAl. The volume shown contains three unit cells and nine U atoms. (b) U-Rh$_{\rm {I}}$ and (c) Rh$_{\rm {II}}$-Al layers viewed along the $c$-axis.}
\end{figure} 

 URhAl orders ferromagnetically at $T_{\rm C}$ = 26 $\sim$ 27 K\cite{veenhuizen,javorsky2}. The value of the linear specific heat coefficient $\gamma$ is 60 mJ/molK$^2$ and the spontaneous magnetic moment is about 1 ${\mu}_{\rm B}$ per uranium ion. Neutron scattering studies show that the magnetic moments point parallel to the $c$-axis and that a magnetic moment is induced only at the Rh I site by the ferromagnetic ordering of the $5f$ moment of the uranium atoms in the same plane\cite{paixao1}. 
 
 The magnetic properties are highly anisotropic with the easy axis parallel to the $c$-axis. The anisotropy could be explained by the crystalline electric field (CEF) effect on the $5f$ electrons but no clear CEF excitation was detected by inelastic neutron scattering\cite{hiess}. None of the bulk physical quantities such as specific heat and magnetization can be easily explained with the CEF model. Rather, it seems to be reasonable to consider itinerancy of the $5f$ electrons in URhAl as suggested from electronic structure calculations\cite{kunes,antonov}. Recently, we have analyzed the magnetic data of 80 actinide ferromagnets using spin fluctuation theory\cite{tateiwa2}. A parameter ${T_{\rm C}}/{T_{\rm 0}}$ indicates the itineracy of magnetic electrons in the theory. Here, ${T_{\rm 0}}$ is the width of the spin fluctuation spectrum in energy space. The magnetic electrons have a strongly itinerant character for ${T_{\rm C}}/{T_{\rm 0}}{\,}{\ll}{\,}1$ but local magnetic moment character for the ferromagnetism when ${T_{\rm C}}/{T_{\rm 0}}{\,}{=}{\,}1$. The value of ${T_{\rm C}}/{T_{\rm 0}}$ is 0.365 in URhAl. The ferromagnetic state in URhAl is located in an intermediate range between itinerant and localized electrons models.

\section{EXPERIMENT and ANALYSIS}
We have grown a high-quality single crystal sample of URhAl by Czochralski pulling in a tetra arc furnace. Magnetization was measured in a commercial superconducting quantum interference (SQUID) magnetometer (MPMS, Quantum Design). We measured a rectangular-shaped single crystal sample and the size of the sample was 0.40 $\times$ 0.30 $\times$ 0.34 mm$^3$. We determined the internal magnetic field ${{\mu}_0}H$ by subtracting the demagnetization field $DM$ from the applied magnetic field ${{\mu}_0}H_{ext}$: ${{\mu}_0}H$ = ${{\mu}_0}H_{ext}$ - $DM$. The demagnetizing factor $D$ was calculated from the macroscopic dimensions of the sample. The critical exponents have been determined using a modified Arrott plot, critical isotherm analysis, a Kouvel-Fisher plot, and scaling analysis. The obtained exponents have been analyzed with a renormalized group approach.  

\section{RESULTS AND DISCUSSIONS}
      \begin{figure}[t]
\includegraphics[width=8.5cm]{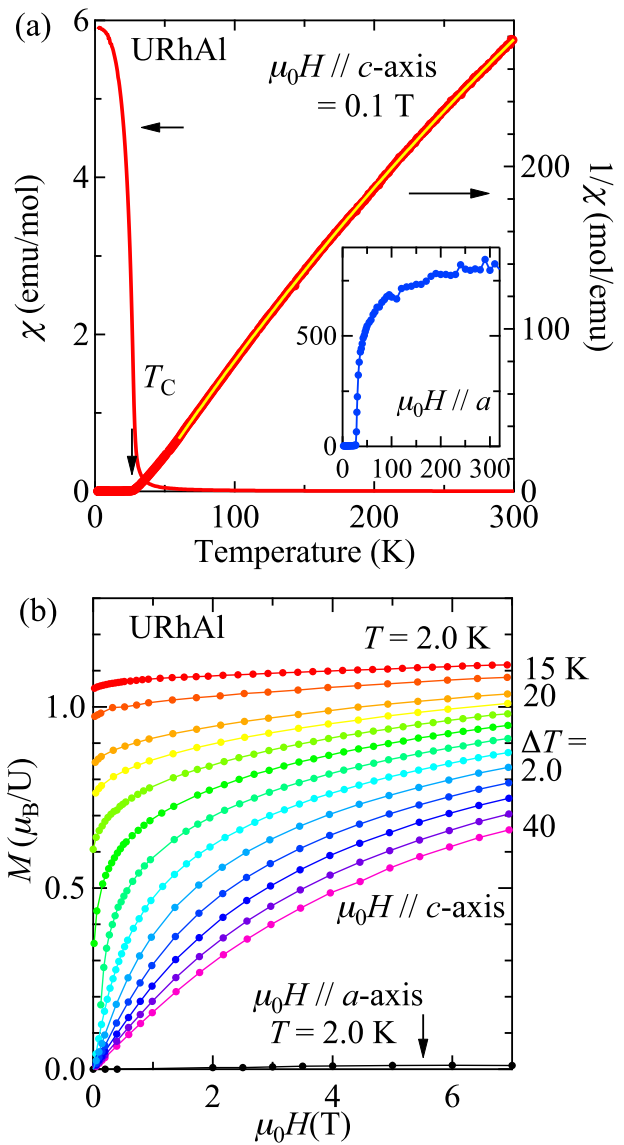}
\caption{\label{fig:epsart}(a)Temperature dependence of the inverse of the magnetic susceptibility $1/{\chi}$ in magnetic field of 0.1 T applied along the magnetic easy $c$-axis in URhAl. Solid line is the result of the fit to the data using a modified Curie-Weiss law. The inset shows the temperature dependence of $1/{\chi}$ in magnetic field applied along the $a$-axis. (b) Magnetic field dependence of the magnetization at several temperatures in magnetic field along the $c$- and $a$-axes in URhAl. }
\end{figure} 

    \begin{figure*}[t]
\includegraphics[width=18cm]{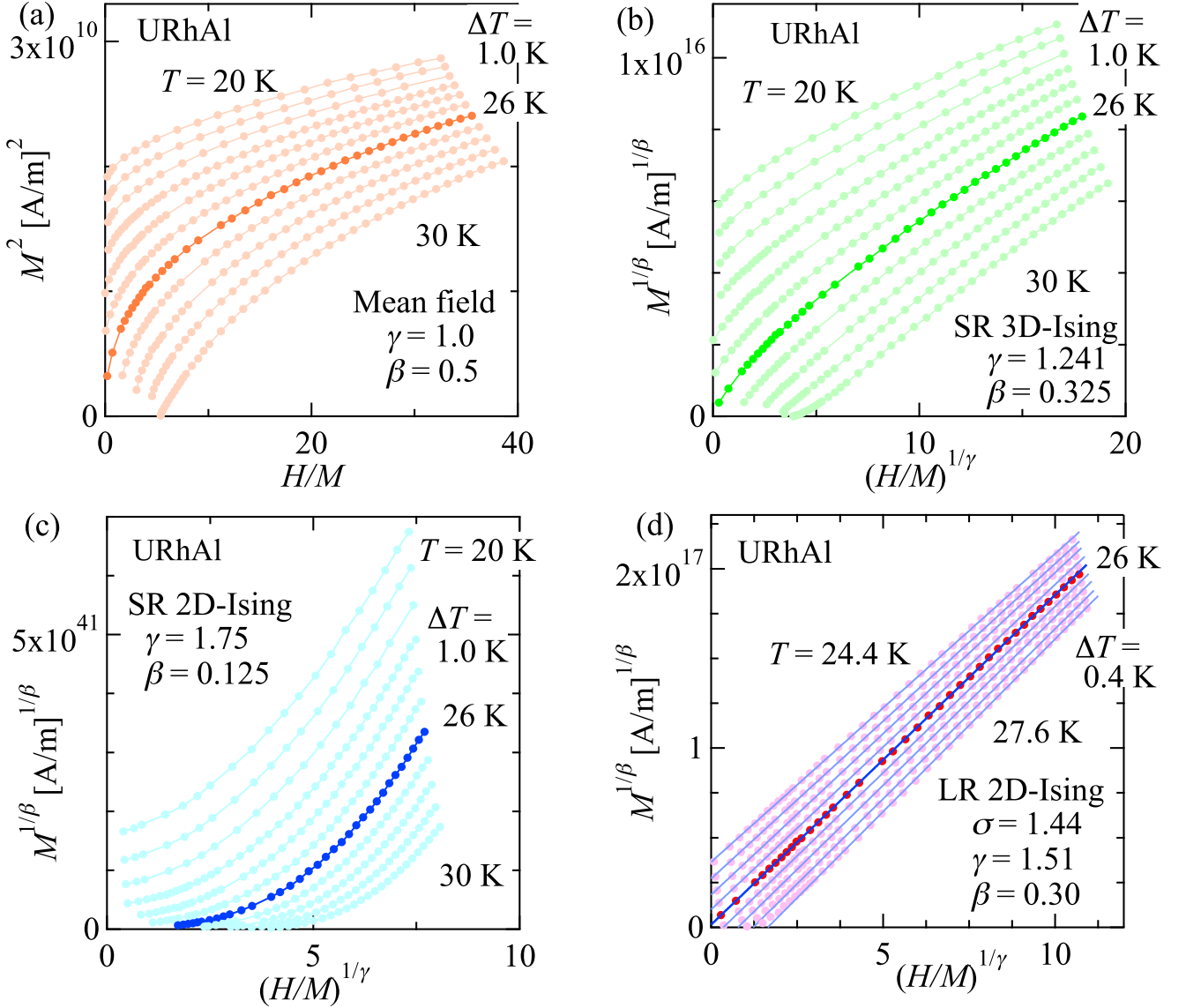}
\caption{\label{fig:epsart}Isotherms of $M^{1/{\beta}}$ vs. $(H/M)^{1/{\gamma}}$ for 20 K $\le$ $T$ $\le$ 30 K, with (a) the mean field theory, (b) the short-range (SR) 3D-Ising model, (c) the SR 2D-Ising model, and (d) the modified Arrott plot of isotherms with ${\beta}$ = 0.30 and ${\gamma}$ = 1.51 in URhAl. Lines in (d) show fits to the data with the equation (6).}
\end{figure*} 

 Figure 2 (a) shows temperature dependence of the magnetic susceptibility ${\chi}$ and its inverse $1/{\chi}$ in a magnetic field of 0.1 T applied along the magnetic easy $c$-axis in URhAl. Solid line is a fit to the data using a modified Curie-Weiss law ${\chi}$ = ${C/(T-{\theta})}+ {{\chi}_0}$. Here, $C$ is the Curie constant, $\theta$ is the paramagnetic Curie temperature, and $ {{\chi}_0}$ is the temperature-independent term that may arise from the density of states at the Fermi energy from other than the $5f$ electrons. The effective magnetic moment, $p_{\rm eff}$, per a magnetic atom is estimated as $p_{\rm eff}$ = 2.50 ${\mu}_{\rm B}$/U from $C$ = ${N_{\rm A}}{{\mu}_{\rm B}^2}{p_{\rm eff}^2}/3{k_{\rm B}}$, where $N_{\rm A}$ is the Avogadro constant. The values of $p_{\rm eff}$ is smaller than that expected for $5f^2$ (U$^{4+}$, $p_{\rm eff}$ = 3.58 ${\mu}_{\rm B}$/U) or $5f^3$ (U$^{3+}$, $p_{\rm eff}$ = 3.62 ${\mu}_{\rm B}$/U) configurations, suggesting itinerant character of the $5f$ electrons. The inset of Fig. 2 (a) shows the temperature dependence of the inverse of the magnetic susceptibility $1/{\chi}$ in a magnetic field of 0.1 T applied along the magnetic hard $a$-axis in URhAl. The magnetic susceptibility shows clear anisotropic behavior in the paramagnetic state.
 
 Fig. 2 (b) shows magnetic field dependencies of the magnetization at several temperatures in magnetic fields applied along the $c$- and $a$-axes in URhAl. The spontaneous magnetic moment $p_{\rm s}$ is determined as $p_{\rm s}$ = 1.05 ${\mu}_{\rm B}$/U from the magnetization curve at 2.0 K for magnetic field along $c$-axis. Clearly, the magnetic anisotropy is huge also in the ferromagnetic ordered state. These basic magnetic properties are consistent with those in the previous studies\cite{sechovsky,veenhuizen}.

    \begin{figure}[]
\includegraphics[width=8.5cm]{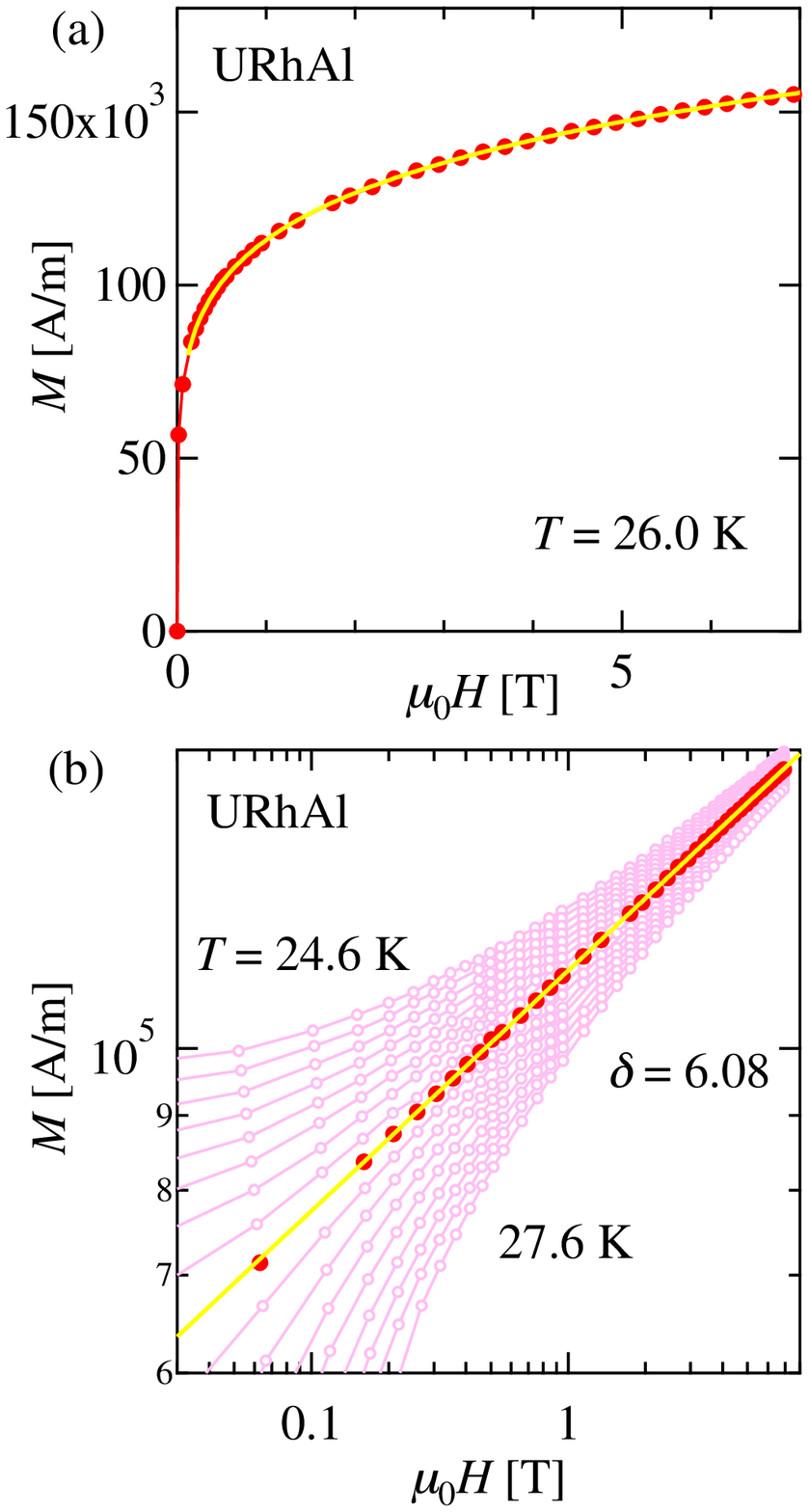}
\caption{\label{fig:epsart}Magnetic field dependencies of the magnetization (a) at 26.0 K, and (b) from 24.6 K to 27.6 K in URhAl. Lines show fit to the isotherm at 26.0 K with the eq. (5) to obtain the critical exponent $\delta$.}
\end{figure} 
   \begin{figure}[]
\includegraphics[width=8.5cm]{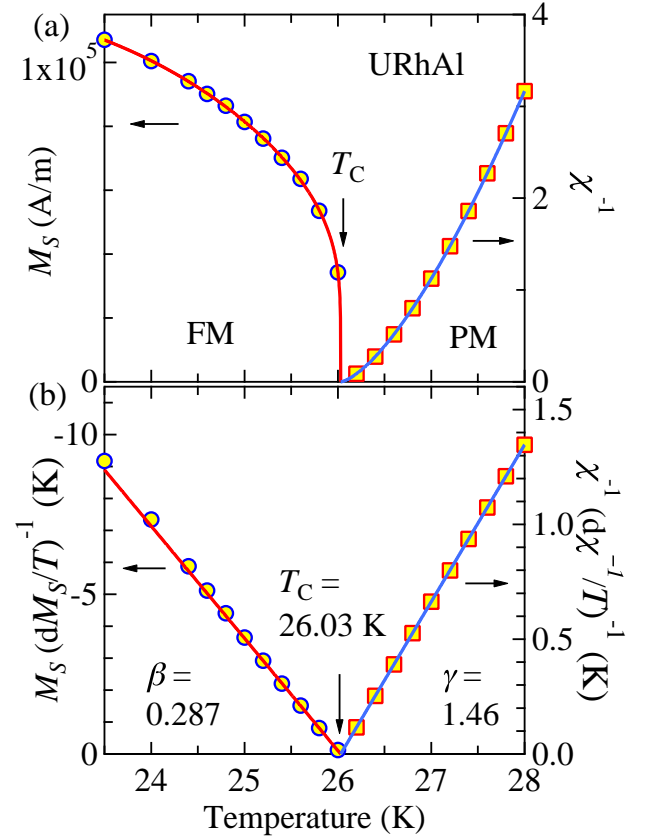}
\caption{\label{fig:epsart}(a) Temperature dependencies of the spontaneous magnetization $M{_s}(T)$ in the ferromagnetic (FM) state and the inverse of the initial magnetic susceptibility ${\chi}^{-1}$ in the paramagnetic state (PM) determined from the modified Arrott plot. (b)Kouvel-Fisher plots for $M{_s}(T)$ and ${\chi}^{-1}$ in URhAl.}
\end{figure} 
        \begin{figure}[]
\includegraphics[width=8.5cm]{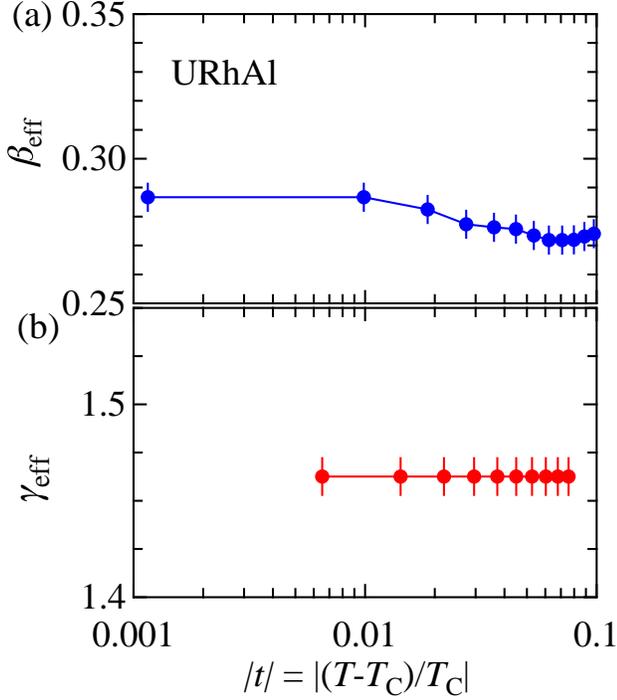}
\caption{\label{fig:epsart}Effective exponents (a) ${\beta}_{\rm eff}$ below $T_{\rm C}$ and (b) ${\gamma}_{\rm eff}$ above $T_{\rm C}$ as a function of reduced temperature $|t|$ (=$|({T}-{T_{\rm C}})/{T_{\rm C}}|$) in URhAl.}
\end{figure} 
    \begin{figure}[t]
\includegraphics[width=8.5cm]{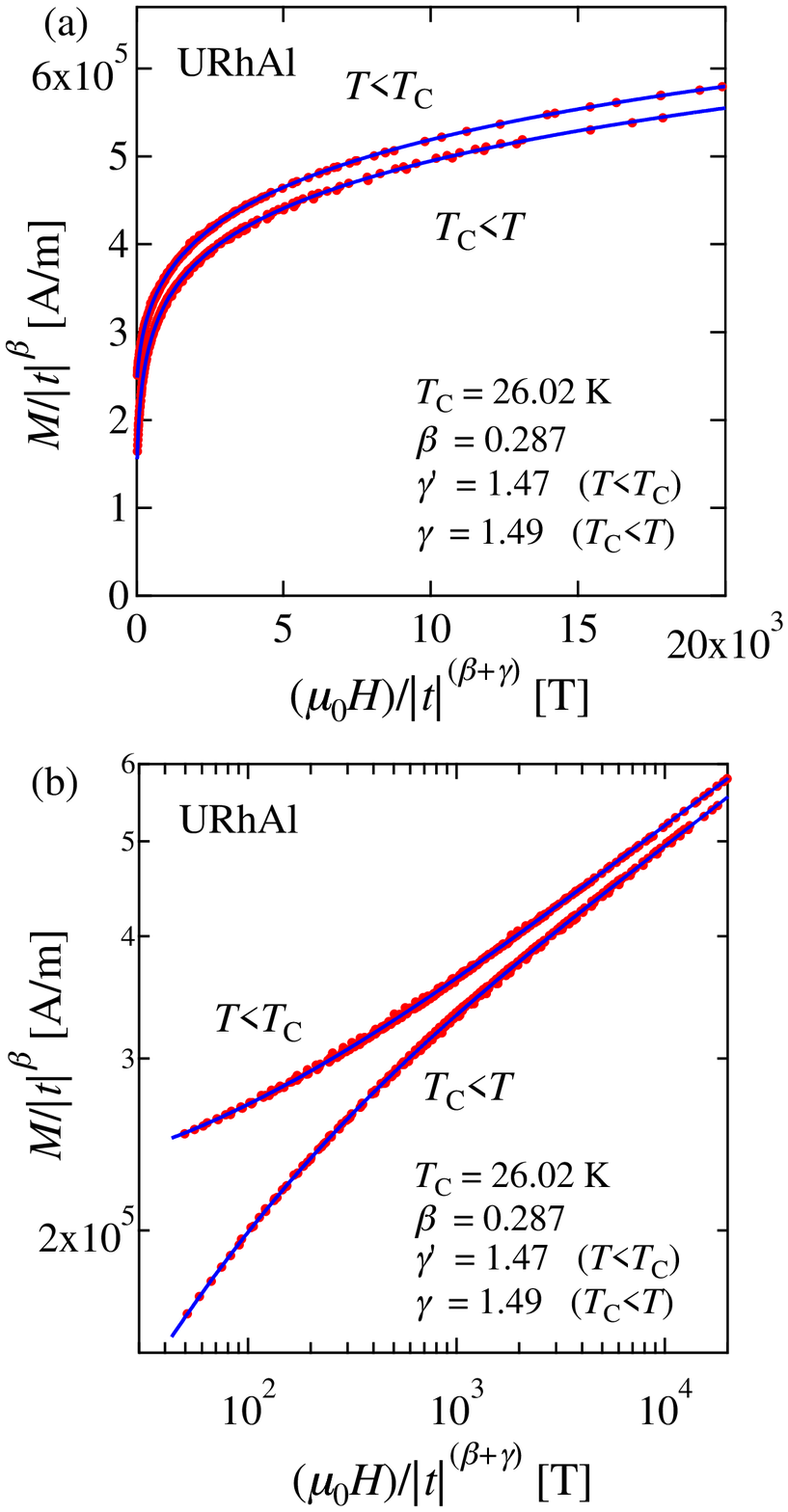}
\caption{\label{fig:epsart}Scaled magnetization as a function of renormalized field following Eq. (11) below and above the critical temperature $T_{\rm C}$ 
for URhAl. Solid lines show best fit polynomials. The magnetization data in the temperature range $t{\,}={\,}|({T}-{T_{\rm C}})/{T_{\rm C}}|{\,}< 0.07$ are shown.}
\end{figure}

  In the Landau (mean field) theory, the free energy of a ferromagnet ${F_m}(M)$ can be expanded as a power series in the order parameter $M$ in the vicinity of a second order phase temperature.
     \begin{eqnarray}
 F{_m}(M)&=&F{_m}(0)+ {1\over 2}aM^2+{1\over 4}bM^4 +....-HM 
  \end{eqnarray}
 The equilibrium condition is obtained from minimizing the thermodynamic potential ${\partial F{_m}}(M)$/${\partial M}$ = 0. The following equation of state is derived for the behavior of the magnetization near the transition temperature. 
      \begin{eqnarray}
H&=&aM+bM^3 
  \end{eqnarray}
 This mean field formula fails in a critical region characterized with the Ginzburg criterion\cite{ginzburg2}. The divergence of correlation length $\xi$ = ${\xi}_0$ $|1-T/{T_{\rm C}}|^{-{\nu}}$ leads to universal scaling laws for spontaneous magnetization $M_S$ and initial susceptibility ${\chi}$ in the critical region. ${{\nu}}$ is the critical exponent. The definitions of exponents are as follows\cite{privman}.   
      \begin{eqnarray}
   {{\chi}}(T){^{-1}}&{\propto}&  {|t|}^{-{{\gamma}}'} {\;}{\;}{\;}{\;}(T<T{_{\rm C}}), {\;}{\;} {|t|}^{-{{\gamma}}} {\;}{\;}(T{_{\rm C}}< T)\\ 
 M{_S}(T) &{\propto}& |t|{^{\beta}}  {\;}  {\;} {\;}{\;}{\;} {\;}{\;}(T < T{_{\rm C}})\\ 
 {M{_S}}& {\propto} & ({{\mu}_0}H)^{1/{\delta}}   {\;} {\;} {\;}{\;}{\;} (T = T{_{\rm C}})
   \end{eqnarray}        
 Here, $t$ denotes the reduced temperature $t$ = $1-{T}/{T_{\rm C}}$. Parameters $\beta$, $\gamma$, ${\gamma}'$  and $\delta$ are the critical exponents. 
 
 The critical exponents and the phase transition temperature $T{_{\rm C}}$ can be determined using Arrott plots. These plots in the form of $M^2$ vs. $H/M$ should show a set of parallel straight lines and the isotherm at $T_{\rm C}$ should pass through origin\cite{privman}. The Arrott plots assume the critical exponents following mean-field theory with $\beta$ = 0.5, $\gamma$ = 1.0, and $\delta$ = 3.0. The $H/M$ vs. $M^2$ plots in URhAl shown in Fig. 3 (a) do not yield straight lines around $T_{\rm C}$, indicating that the mean field model is not valid. Neither the 3D Ising model ($\beta$ = 0.325, $\gamma$ = 1.241) nor 2D Ising model ($\beta$ = 0.125, $\gamma$ = 1.75) with short-range (SR) exchange interactions is appropriate to describe the critical behavior of the magnetization shown in Fig. 3 (b) and (c). The two plots do not exhibit straight lines. Therefore, the magnetization isotherms have been re-analyzed with the Arrott-Noakes equation of state which should hold in the asymptotic critical region\cite{arrott}.

   \begin{eqnarray}
  &&(H/M){^{1/{\gamma}}} = (T-{T_{\rm C}})/{T_1} + (M/{M_{1}})^{1/{\beta}}
   \end{eqnarray}
, where $T_1$ and $M_1$ are material constants. In the corresponding modified Arrott plots, the data for URhAl are represented in the form of $M^{1/{\beta}}$ versus $(H/M)^{1/{\gamma}}$. Then, we have chosen the values of $\beta$ and $\gamma$ in such a way that the isotherms display as closely as possible a linear behavior as shown in Fig. 3 (d). A best fit of equation (6) to the data in URhAl for 24.4 K $\le$ $T$ $\le$ 27.6 K and 0.1 T $\le$ ${{\mu}_0}H$ $\le$ 7.0 T yields $T_{\rm C}$ = 26.05 $\pm$ 0.05 K, $\beta$ = 0.300 $\pm$ 0.002, and ${\gamma}$ = 1.51 $\pm$ 0.02. The obtained critical exponents are shown in Table. I. 

   We determine the third critical exponent $\delta$ from the critical isotherm at $T_{\rm C}$ according to the eq (5) as shown in Figure 4. From fits to the isotherms at 26.0 K, the value of $\delta$ was obtained as $\delta$ = 6.08 $\pm$ 0.02 for URhAl. The value is larger than that in the 3D Ising model with short-range exchange interactions ($\delta$ = 4.80). According to the Widom scaling law, the exponents $\delta$, $\gamma$, and $\delta$ should satisfy the relation ${\delta}$ = 1+${\gamma}/{\beta}$\cite{widom}. The value of ${\delta}$ was estimated as 6.03 $\pm$ 0.10 from the values of $\beta$ and $\gamma$ determined in the modified Arrott plots using the relation. This result is consistent with that determined from the critical isotherm. 

 Next, we analyze the data using the Kouvel-Fisher (KF) method by which the exponents $\beta$ and $\gamma$ can be determined more accurately\cite{kouvel}. The spontaneous magnetization $M{_s}$ is determined from the intersection of the $M^{1/{\beta}}$-axis and the straight lines at the value $M{_s}^{1/{\beta}}$, and ${\chi}^{-1}$ is determined from that of the $(H/M)^{1/{\gamma}}$ axis and the lines at ${\chi}^{-1/{\gamma}}$ in the Arrott plots. Figure 5 (a) shows the temperature dependencies of the spontaneous magnetization $M{_s}$ and the initial magnetic susceptibility ${\chi}$ in URhAl. Solid lines represent fits to the data using Eqs. (3) and (4) for ${\chi}^{-1}(T)$ and $M{_s}(T)$, respectively. The KF method is based on following two equations:
 
        \begin{eqnarray}
 M{_S}(T)[dM{_S}(T)/dT]{^{-1}} &=& (T-T{_{\rm C}}^{-})/{\beta}(T)\\ 
  {{\chi}}{^{-1}}(T)[d {{\chi}}{^{-1}}(T)/dT]{^{-1}} &=& (T-T{_{\rm C}}^{+})/{\gamma}(T)
   \end{eqnarray}

 Eq. (6) can be reduced to Eqs. (7) and (8) in the limit $H$ $\rightarrow$ 0 for $T$ $<$ and $>$ $T_{\rm C}$, respectively. The quantities ${\beta}(T)$ and ${\gamma}(T)$ become identical with the critical values $\beta$ and $\gamma$, respectively, in the limit $T$ $\rightarrow$ $T_{\rm C}$. We can determine the values of $\beta$ and $\gamma$ from the slope of $M{_s}(T)[dM{_S}(T)/dT]{^{-1}}$ and $ {{\chi}}{^{-1}}(T)[d {{\chi}}{^{-1}}(T)/dT]{^{-1}}$-plots, respectively, at $T_{\rm C}$ and the intersection with the $T$-axis yields $T_{\rm C}$ as shown in Fig. 5 (b). Solid lines represent the fits to the data using Eqs. (7) and (8). The exponents $\beta$ and $\gamma$ for URhAl are determined as $\beta$ = 0.287 $\pm$ 0.002 and $\gamma$ = 1.46 $\pm$ 0.03 with $T_{\rm C}$ = ($T{_{\rm C}}^{+}$ + $T{_{\rm C}}^{-}$)/2 = 26.03 $\pm$ 0.02 K by the KF method. The results are consistent with those determined in the modified Arrott plot.

     \begin{table*}[]
\caption{\label{tab:table1}%
Comparison of critical exponents $\beta$, $\gamma$, and $\delta$ of URhAl with various theoretical models. Abbreviations; RG-${\phi}^4$: renormalization group ${\phi}^4$ field theory,  RG-${\epsilon}'$: renormalization group epsilon (${\epsilon}'=2{\sigma}-d$) expansion, SR: short-range, LR: long-range.}
\begin{ruledtabular}
\begin{tabular}{ccccccccc}
\textrm{}&
\textrm{Method}&
\textrm{$T{_{\rm C}}$}&
\textrm{$\beta$ }&
\textrm{${\gamma}'$}&
\textrm{${\gamma}$}&
\textrm{$\delta$}&
\textrm{Reference}&\\
\textrm{}&
\textrm{}&
\textrm{}&
\textrm{}&
\textrm{($T<T{_{\rm C}}$)}&
\textrm{($T{_{\rm C}}<T$)}&
\textrm{}&
\textrm{}&\\
\colrule
(Theory)&&&&&&&&\\ 
Mean field &&&0.5&\multicolumn{2}{c}{1.0}&3.0&&\\ 
SR exchange: $J(r){\,}{\sim}{\,}e^{-r/b}$ &&&&&&&&\\ 
$d$ =  2, $n$ =1 &Onsager solution&&0.125&\multicolumn{2}{c}{1.75}&15.0&\cite{privman,fischer1}&\\ 
$d$ =  3, $n$ =1 &RG-${\phi}^4$&&0.325&\multicolumn{2}{c}{1.241}&4.82&\cite{privman,guillou}&\\ 
$d$ =  3, $n$ =2 &RG-${\phi}^4$&&0.346&\multicolumn{2}{c}{1.316}&4.81&\cite{privman,guillou}&\\ 
$d$ =  3, $n$ =3 &RG-${\phi}^4$&&0.365&\multicolumn{2}{c}{1.386}&4.80&\cite{privman,guillou}&\\ 
LR exchange: $J(r){\,}{\sim}{\,}r^{-(d+{\sigma})}$ &&&&&&&&\\ 
$d$ =  2, $n$ =1, $\sigma$ = 1.44  &RG-${\epsilon}'$&&0.289&\multicolumn{2}{c}{1.49}&6.16&\cite{fischer1}&\\ 
\colrule

(Experiment)&&&&&&&&\\ 
URhAl&Modified Arrott &26.05 $\pm$ 0.05  &0.300 $\pm$ 0.002 &  \multicolumn{2}{c}{1.51 $\pm$ 0.02} &&This work&\\ 
&Kouvel-Fisher &26.03  $\pm$ 0.02  &0.287 $\pm$ 0.002 &  \multicolumn{2}{c}{1.46 $\pm$ 0.03}  &&&\\
&Scaling &26.02  $\pm$ 0.02  &0.287 $\pm$ 0.005& 1.47 $\pm$ 0.02 & 1.49 $\pm$ 0.02 &&& \\
&ln$(M)$ vs. ln${({{\mu}_0}H)}$ &  & & && 6.08 $\pm$ 0.04 && \\
\end{tabular}
\end{ruledtabular}
 \end{table*}
 Various systematic trends or crossover phenomena in the critical exponents could appear on approaching $T_{\rm C}$ when a magnetic system is governed by various competing couplings or disorders. To check this possibility, we obtain effective exponents ${\beta}_{\rm eff}$ and ${\gamma}_{\rm eff}$ as follows. 

   \begin{eqnarray}
{{\beta}_{\rm eff}} (t) =  d[{\rm ln}M{_s}(t)]/d({{\rm ln}{t}}),\\{{\gamma}_{\rm eff}} (t) =  d[{\rm ln}{{\chi}^{-1}}(t)]/d({{\rm ln}{t}})
   \end{eqnarray}

Figure 6 (a) and (b) show the effective exponents ${\beta}_{\rm eff}$ and ${\gamma}_{\rm eff}$ as a function of reduced temperature $t$ in URhAl. The exponents ${\beta}_{\rm eff}$ and ${\gamma}_{\rm eff}$ show a monotonic $|t|$-dependence for ${|t|}{\,}{\geq}$ 1.15${\times}{10^{-3}}$ and 6.53${\times}{10^{-3}}$, respectively. This rules out the possibility that the obtained exponents happen to appear around a crossover region between two universality classes as reported in Ni$_3$Al\cite{semwal}.
 
 It is necessary to check whether the set of the critical exponents are the same below and above $T_{\rm C}$. We can determine separately the values of ${\gamma}$' ($T<T{_{\rm C}}$) and $\gamma$ ($T{_{\rm C}}<T$) with scaling theory that predicts the existence of a reduced equation of state close to the ferromagnetic transition temperature\cite{privman}: 

    \begin{eqnarray}
  &&M({{\mu}_0}H, t) = {|t|^{\beta}} f{_{\pm}}({{\mu}_0}H/|t|^{{\beta}+{\gamma}})
   \end{eqnarray}
, where $f_{+}$ for $T{_{\rm C}} < T$ and $f_{-}$ for $T < T{_{\rm C}}$ are regular analytical functions. We can rewrite the scaling equation as $m$ = $f{_{\pm}}{(h)}$ with the renormalized magnetization $m$ and the renormalized field $h$ defined as $m$ $\equiv$ ${|t|^{-{\beta}}}{M({{\mu}_0}H, t)}$ and $h$ $\equiv$ ${H}{|t|^{-({\beta}+{\gamma})}}$, respectively. When the correct $\beta$, $\gamma$, and $t$ values are chosen, the data points in the plot of $M({{\mu}_0}H, t)/{|t|^{\beta}}$ versus ${{\mu}_0}H/|t|^{{\beta}+{\gamma}}$ should fall on two universal curves: one for $T < T{_{\rm C}}$ and the other for $T > T{_{\rm C}}$. The scaled magnetization as a function of renormalized field below and above $T_{\rm C}$ in URhAl is shown in Figure 7 (a) and (b). We show the magnetization data in the temperature range $t{\,}={\,}|({T}-{T_{\rm C}})/{T_{\rm C}}|{\,}< 0.07$. All data points fall on two curves when $T_{\rm C}$ and the critical exponents are chosen as $T_{\rm C}$ = 26.02 $\pm$ 0.02 K, $\beta$ = 0.287 $\pm$ 0.005, ${\gamma}'$ = 1.47 $\pm$ 0.02 for $T$ $<$ $T_{\rm C}$, and $\gamma$ = 1.49 $\pm$ 0.02 for $T_{\rm C}$ $<$ $T$ in URhAl.

  Table I shows the critical exponents $\beta$, $\gamma$, and $\delta$ determined for URhAl and theoretical ones for various models\cite{privman,fischer1,guillou}. The obtained critical exponents in URhAl differ from those of the 3D Heisenberg ($d$ =  3, $n$ =3), 3D XY ($d$ =  3, $n$ =2), 3D Ising ($d$ =  3, $n$ =1) models and 2D Ising ($d$ =  2, $n$ =1) models with short-range (SR) exchange interactions $J(r){\,}{\sim}{\,}e^{-r/b}$, where $b$ is the correlation length. The values of $\beta$ are smaller and the $\gamma$ and $\delta$ values are larger than those of the 3D models. The exponents in URhAl are significantly different from those in the mean field theory and the 2D Ising model. Several reasons can be considered for differences between the critical exponents in real magnets and the theoretical ones as we have discussed for unconventional critical phenomena in UGe$_2$ and URhGe\cite{tateiwa1}. Here, we propose that the long-range nature of magnetic exchange interactions plays an important role in the critical phenomenon in URhAl. 
 
  In the theoretical models with short-range interactions, the interaction between the magnetic moments falls off rapidly with distance. However, the interaction can be of long-range due to mobile electrons for the itinerant electron system. The universality class of the magnetic phase transition depends on the range of the exchange interaction $J(r)$. The fixed point of a system with short-range exchange interactions becomes unstable due to long-range interactions, which leads to a crossover to the fixed point with long-range interaction. The critical exponents are shifted towards those of the mean field theory. A renormalization group theory analysis has been done by Fischer {\it et al.} for systems with the magnetic exchange interaction of a form $J(r)$ $\sim$ $1/r^{d+{\sigma}}$, where $d$ is the dimension of the system and $\sigma$ is the range of exchange interaction\cite{fischer2}. The analysis showed the validity of such a model with long-range interactions for $\sigma$ $<$ 2. The critical exponents in ferromagnetic nickel are slightly shifted from those of the 3D Heisenberg ($d$ =  3, $n$ = 3) model with short-range interactions towards the mean field values and the deviations can be understood with the renormalization group theory analysis\cite{seeger}. The exponent $\gamma$ in the theory is expressed as follows.
     \begin{figure}[]
\includegraphics[width=8.5cm]{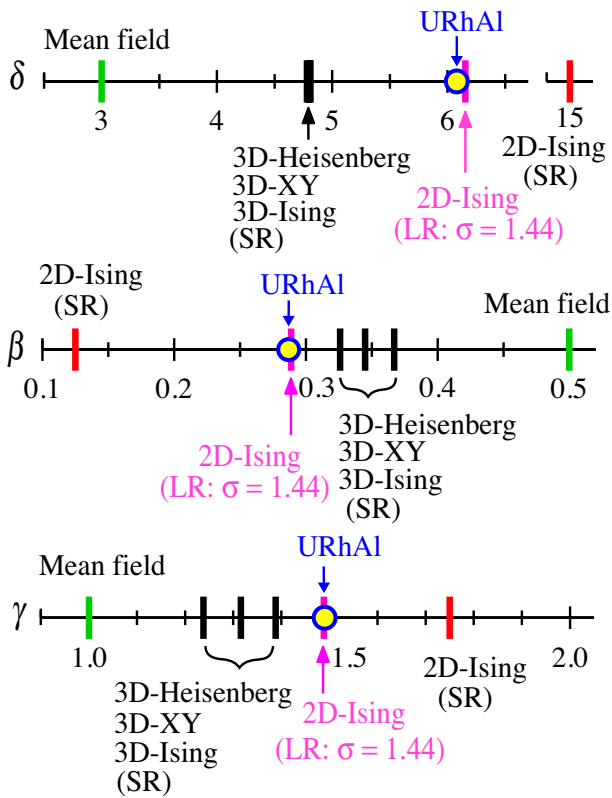}
\caption{\label{fig:epsart}
Comparison of the critical exponents $\beta$, $\gamma$, and ${\delta}$ in URhAl denoted as closed circles with those of known universality classes shown as vertical bars: Mean field, 2D-Ising, 3D-Ising, 3D-XY, and 3D-Heisenberg models with short range exchange interactions, and 2D-Ising model with long-range ($\sigma$ = 1.44) interactions.}
\end{figure}

  \begin{eqnarray}
{\gamma} = 1+{4\over d}{\biggl(}{{{n+2}\over {n+8}}}{\biggl)}{{\Delta}{\sigma}}+{{8(n+2)(n-4)}\over {{d^2}{(n+8)^2}}}\nonumber\\
{\times}{\Biggl[}1+{{2G({d\over 2})(7n+20)}\over {{(n-4)(n+8)}}}{\Biggl]}{{\Delta}{\sigma}^2}
 \end{eqnarray} 
, where ${{\Delta}{\sigma}}{\,}={\,}{\sigma}-{d\over 2}$, $G({d\over 2}){\,}={\,}3-{1\over 4}({d\over 2})^2$, and $n$ is the spin dimensionality. This expression holds for ${d/2}{\,}{\le}{\,}{\sigma}{\,}{\le}{\,}2$. The theoretical models with short-range interaction valid for $2{\,}{<}{\sigma}$ and the mean field model describes the critical behavior for ${\sigma}{\,}{<}{\,}{d/2}$.

 We have examined this theory for the critical exponents in URhAl. The parameter $\sigma$ was chosen for a particular set of ${\{}d:n{\}}$ in such that Eq. (12) for $\gamma$ yields a value close to that  ($\sim$ 1.5) determined experimentally. We obtained the other exponents $\alpha$, $\beta$ and $\delta$ using scaling relations $\alpha$ = $2-{\nu}d$, $\beta$ = $(2-{\alpha}-{\gamma})/2$ and $\delta$ = 1+ $\gamma$/$\beta$ where $\eta$ = $2-{\sigma}$ and $\nu$ = ${\gamma}/{\sigma}$. The best match to the obtained critical exponents is obtained for $d$ = 2, $n$ = 1, and $\sigma$ = 1.44 after repeating this procedure for different sets of ${\{}d:n{\}}$ ($d$, $n$ = 1, 2, 3). The long-range 2D Ising model with $\sigma$ = 1.44 produces the critical exponents $\beta$ = 0.289, $\gamma$ = 1.49, and $\delta$ = 6.15. These exponents match very well with the obtained result in URhAl as shown in Fig. 8. The obtained critical exponents in URhAl are located between those of the 2D Ising model with short-range interactions and the mean field theory.

We have examined other 2D and 3D models but failed to explain the experimental result. For example, the value of the exponent $\gamma$ is 1.386, 1.316, and 1.241 for the 3D Heisenberg, XY and Ising models with short-range exchange interactions ($2{\,}{<}{\sigma}$). When the range of the interaction becomes longer, that is, ${\sigma}$ decreases from 2, the $\gamma$-values decrease and approach to the mean field value (${\gamma}=$1.0) for ${\sigma}{\rightarrow}{\,}{d/2}= 3/2$. There is no choice of $\sigma$ in the permissible range $3/2{\,}{\le}{\,}{\sigma}{\,}{\le}{\,}2$ when the $\gamma$-value is substituted into Eq. (12). The obtained critical exponents URhAl cannot be explained by the 3D models with long-range interactions. Next, we examined the 2D XY model. The reasonable value of $\sigma$ = 1.56 was obtained using Eq. (12) with the $\gamma$-value. However, the calculated values of the other exponents ($\beta$ = 0.213 and $\delta$ = 8.04) do not match with those determined experimentally ($\beta$ = 0.287 $\sim$ 0.300, $\delta$ = 6.08). The ferromagnetic state in URhAl has a strong uniaxial magnetic property. The magnetic moments point along the $c$-axis: perpendicular to the U-Rh$_{\rm I}$ layer. The magnetic exchange interaction between the uranium ions in the layer was suggested in the neutron scattering experiment\cite{paixao1}. The 2D Heisenberg and XY models, and the 1D models are not suitable. We conclude that the 2D Ising model with long-range interactions is appropriate to explain the critical behavior of the magnetization in URhAl. 

 As mentioned in the introduction, we have reported the unconventional critical scaling in uranium ferromagnetic superconductors UGe$_2$ and URhGe\cite{tateiwa1}. This unconventional critical scaling of the magnetization cannot be explained by the present renormalization group theory analysis. It was theoretically analyzed with a non-local Ginzburg-Landau model where the quartic nonlocality arises as a result of magnetoelastic interaction\cite{singh1}.

  We discuss the low dimensionality in the magnetic properties of URhAl. As shown in Figure 1 (a)-(c), the constituting atoms are arranged in planer layers perpendicular to the $c$-axis . The layer of the U and Rh I sites alternates with that of the Rh II and Al sites.  As mentioned in the introduction, a magnetic magnetic moment of 0.28 ${\mu}_{\it B}$ is induced at the Rh I site below $T_{\rm C}$ but the moment of the Rh II site at the adjacent layer is zero within statistical accuracy\cite{paixao1}. This suggests the strong anisotropic hybridization between the $5f$ electrons of the U atom and the $4d$ electrons of the Rh atom in the same plane. This peculiar hybridization inside the U-Rh$_{\rm I}$ layer may give rise to the two dimensional character seen in the magnetic property of URhAl. 
      
  We compare URhAl with UCoAl. As mentioned in the introduction, experimental data on UCoAl have been discussed from the viewpoints of quantum criticality based on the 3D Ising system\cite{aoki,matsuda2,nohara,karube1,matsuda3,morales,combier1,takeda,karube2,shimizu1,kimura,combier2}. It was concluded that the metamagnetic transition at the critical end-point belongs to the 3D Ising universality class\cite{karube1}. Polarized neutron diffraction studies have shown that the magnetic moments of both the Co I and Co II sites are induced but the magnitude of the induced moments on the Co sites are less than 20{\%} of that at the Rh$_{\rm I}$ site in URhAl\cite{javorsky1,papoular}. This difference could be ascribed to the weaker and more isotropic hybridization between the $3d$ electrons of the Co atom and the $5f$ electrons in UCoAl. The wave functions of the $4d$ states in the Rh atom are more expanded in real space than those of the $3d$ states in the Co atom. The strong and anisotropic hybridization between the $5f$ electrons in the U atom and $4d$ electrons in the Rh$_{\rm I}$ atoms may be determining the low dimensionality of the critical phenomenon in URhAl.

 Ref. 21 reported the pressure-temperature-magnetic field phase diagram in URhAl determined by the electrical resistivity measurement\cite{shimizu2}. The application of the pressure induces a ferromagnetic to  non-magnetic transition at critical pressure $P_{\rm c}$ $\sim$ 5.2 GPa and non Fermi liquid behavior in the resistivity is observed above $P_{\rm c}$. The pressure effect on the electrical resistivity ${\rho}(T)$ at low temperatures was analyzed with a form ${\rho}(T)={{\rho}_0}+{A^{'}}T^n$. ${{\rho}_0}$ is the residual resistivity. The resistivity exponent $n$ just above $P_{\rm c}$ is 1.6 $\sim$ 1.7. This value is close to the exponent $n$ = 5/3 around the phase boundary of the 3D ferromagnetism in the self-consistent renormalization (SCR) theory for spin fluctuations\cite{moriya1}. Note that the exponent $n$ for the 2D ferromagnetism is 4/3 in the SCR theory\cite{hatatani}. The obtained resistivity exponent $n$ in Ref. 21 does not seem to be compatible with the present study. Further theoretical consideration on this discrepancy is necessary. We suggest one possibility that the dimensionality of the ferromagnetism in URhAl changes under high pressure where the crystal is compressed. We also point out the needs for future theoretical study for the behavior of the electrical resistivity around the phase boundary of a magnetic ordered state with long-range exchange interactions. 
 
Finally, we discuss the consequence of the present study. The finding of the ferromagnetic superconductivity in UGe$_2$, URhGe, and UCoGe has triggered extensive studies on the quantum phase transition between the ferromagnetism and paramagnetism induced by the application of high pressure and high magnetic field in uranium ferromagnets. Generally, novel features of the physical properties around the transition have been discussed with theories based on the 3D Ising model. The present study for URhAl and the previous one for the ferromagnetic superconductors UGe$_2$ and URhGe provide different views to this research field. Recently, the low-dimensionality of the magnetism has been extensively studied in several itinerant ferromagnets in the $3d$ electrons systems\cite{bhattacharyya,han,liu,liu2,lin,liu3}, while it has been rarely recognized in studies on uranium intermetallic compounds. We hope the present result prompts further progress for the understanding the quantum phase transition in uranium ferromagnets.

\section{SUMMARY}
We have studied the critical behavior of the magnetization in uranium ferromagnet URhAl at around its ferromagnetic transition temperature $T_{\rm C}$ = 26.02 $\pm$ 0.02 K. The critical exponent $\beta$ for the temperature dependence of the spontaneous magnetization below $T_{\rm C}$, $\gamma$ for the magnetic susceptibility, and $\delta$ for the magnetic isotherm at $T_{\rm C}$ have been determined with a modified Arrott plot, a Kouvel-Fisher plot, the critical isotherm analysis, and the scaling analysis. The critical exponents have been determined as $\beta$ = 0.287 $\pm$ 0.005, ${\gamma}'$ = 1.47 $\pm$ 0.02 for $T$ $<$ $T_{\rm C}$, $\gamma$ = 1.49 $\pm$ 0.02 for $T_{\rm C}$ $<$ $T$, and $\delta$ = 6.08 $\pm$ 0.04 by the scaling analysis and the critical isotherm analysis. The obtained critical exponents satisfy the Widom scaling law ${\delta}{\,}={\,}1+{\,}{\gamma}/{\beta}$. Although uniaxial magnetic properties in URhAl and its iso-structural UCoAl has been discussed based on the 3D Ising model in previous studies, the universality class of the critical phenomenon in URhAl does not belong to the 3D Ising system ($\beta$ = 0.325, $\gamma$ = 1.241, and $\delta$ = 4.82) with short-range exchange interactions between magnetic moments. The determined exponents match well with those calculated from the renormalization group approach for a two-dimensional Ising system coupled with long-range interactions decaying as $J(r){\,}{\sim}{\,}r^{-(d+{\sigma})}$ with $\sigma$ = 1.44. We suggest that the strong hybridization between uranium $5f$ and rhodium $4d$ electrons in the U-Rh$_{\rm I}$ layer in the hexagonal crystal structure takes an important role in the low-dimensionality of the critical phenomenon. The consequence of the present result for studies on uranium ferromagnets is discussed. 
 
 \section{ACKNOWLEDGMENTS}
This work was supported by JSPS KAKENHI Grant Number JP16K05463. We thank Prof. Z. Fisk for discussions and his editing of this paper.

\bibliography{apssamp}

\end{document}